# Enhanced upper critical field in Co-doped Ba122 superconductors by lattice defect tuning

Shinnosuke Tokuta[a)] and Akiyasu Yamamoto

Department of Applied Physics, Tokyo University of Agriculture and Technology, 2-24-16 Nakacho, Koganei, Tokyo 184-8588, Japan

[a)]Email: s195941r@st.go.tuat.ac.jp

**Abstract**

Nanoscale defects in superconductors play a dominant role in enhancing superconducting properties through electron scattering, modulation of coherence length, and correlation with quantized magnetic flux. For iron-based superconductors (IBSCs) that are expected to be employed in high-field magnetic applications, a fundamental question is whether such defects develop an upper critical field ($H_{c2}$) similar to that of conventional BCS-type superconductors. Herein, we report the first demonstration of a significantly improved $H_{c2}$ in a 122-phase IBSC by introducing defects through high-energy milling. Co-doped Ba122 polycrystalline bulk samples (Ba(Fe,Co)$_2$As$_2$) were prepared by sintering powder which was partially mechanically alloyed through high-energy milling. A remarkable increase in full-width at half maximum of X-ray powder diffraction peaks, anomalous shrinkage in the $a$-axis, and elongation in the $c$-axis were observed. When lattice defects are introduced into the grains, semiconductor behavior of the electric resistivity at low temperature ($T < 100$ K), slight decrease in transition temperature ($T_c$), upturn of $H_{c2}(T)$ near $T_c$, and a large increase in $H_{c2}(T)$ slope were observed. The slope of $H_{c2}(T)$ increased approximately by 50%, $i.e.$, from 4 to 6 T/K, and exceeded that of single crystals and thin films. Defect engineering through high-energy milling is expected to facilitate new methods for the designing and tuning of $H_{c2}$ in 122-phase IBSCs.

Iron-based superconductors (IBSCs)[1,2] discovered by Kamihara and Hosono *et al.* have attracted significant attention owing to their unconventional pairing mechanism and unique physical properties. From a chemical perspective, IBSCs have many parent materials, *e.g.*, 1111, 122, and 11, and variations of constituent element choice. For example, in $BaFe_2As_2$ (Ba122), superconductivity can be induced by hole doping via K substitution on the Ba site, electron doping via Co substitution on the Fe site, or chemical pressure by P substitution on the As site. IBSCs are an excellent candidate for strong superconducting magnets owing to their high critical temperature ($T_c$) and upper critical field ($H_{c2}$)[2]. 122-phase IBSCs demonstrate small electromagnetic anisotropy[3-10], high irreversibility field that is close to $H_{c2}$[4], and a critical grain boundary angle that is twice as large as that of yttrium barium copper oxide (YBCO)[11,12], wherein application in polycrystalline form is expected. In fact, a 1-cm diameter compact bulk magnet that traps 1 T[13] and tapes exhibiting transport $J_c$ exceeding $1 \times 10^5$ A/cm$^2$ at 4.2 K and 10 T[14,15] have been developed recently using K-doped Ba122.

Tuning the magnetic phase diagram of a superconducting material is important in both basic and application aspects. Upon comparing 122 to other materials, it is interesting to note that there are very few reports of enhanced $H_{c2}$ via defects such as those caused by particle irradiation[16]. Furthermore, there is no substantial difference in $H_{c2}$ among single crystals, polycrystalline bulks and wires, and thin films. Specifically, in the case of a 8% Co-doped Ba122, $H_{c2}$(0 K) is $\approx 60$ T[4,5,8,17]. In conventional superconductors, $H_{c2}$ can be improved by introducing defects caused by particle irradiation or by adding nonmagnetic impurities. For example, in $Nb_3Sn$, $H_{c2}$(0 K) is improved by fast neutron irradiation[18,19], proton irradiation[20], or ball-milling the elements[21]. In $MgB_2$, $H_{c2}$(0 K) is improved approximately twofold by thermal neutron irradiation[22] and C doping[23,24]. Moreover, in $MgB_2$, $H_{c2}$(0 K) is multiplied several times up to 15–70 T[25,26] from clean single crystals to moderate bulks/wires and dirty thin films. In the case of IBSCs, magnetic scattering by excess Fe in $Fe_{1+y}(Te_{1-x}Se_x)$ increases a slope of $H_{c2}(T)$[27]. Herein, for Ba122 polycrystalline bulks, high-energy ball-milling conditions of a precursor powder were changed systematically and the effects of introducing defects to $H_{c2}$ were evaluated. To facilitate comparison with previous studies, experiments were conducted on Co-doped Ba122 for which single crystal and thin film data are available.

Ba(Fe,Co)$_2$As$_2$ polycrystalline bulk samples were prepared by sintering mechanically alloyed powders. To prevent oxidization, all powder processing was performed in a high purity Ar glove box. Elemental metals, such as Ba, Fe, Co, and As (molar ratio, 1:1.84:0.16:2), were ball-milled using a planetary ball-mill apparatus (Premium line P-7, Fritsch). Herein, the ball-milling condition was varied systematically by changing the ball-milling time and evaluated quantitatively by ball-milling energy ($E_{BM}$), which is an applied energy to powder per mass. $E_{BM}$ is expressed as follows:

$$E_{BM} = c\beta \frac{\left(\omega_p r_p\right)^3}{r_v} t$$

where $c$ is the dimensionless constant of the order of 0.1, $\beta$ is the mass ratio of balls to the powder, $\omega_p$ is the angular frequency, $r_p$ is the revolution radius, $r_v$ is the rotation radius, and $t$ is the ball-milling time[28]. The milled powders were pressed into disk-shaped pellets with a diameter and thickness of 7 and 1.2 mm, respectively. The pellets were vacuum sealed in quartz tubes and sintered at 600°C for 48 h. The obtained polycrystalline bulk Ba(Fe,Co)$_2$As$_2$ samples had relative densities of 65–75%. For reference, a sample synthesized from hand-milled powder mixed in a mortar is denoted as 0 MJ/kg. The phase and structural properties were analyzed by powder X-ray powder diffraction (XRD) (D2 PHASER, Bruker) for the milled powders prior to sintering and ground powders of sintered samples using $CuK\alpha$ ($\lambda = 1.5418$Å) radiation. Lattice parameters $a$ and $c$ were calculated by Rietveld refinements (DIFFRAC.TOPAS). Co-doping levels were estimated by energy-dispersive X-ray spectroscopy (EDS) (QUANTAX, XFlash, Bruker) for the polished surface of the samples. The electrical resistivity measurements were performed under 0–9 T using the conventional four probe method with a physical property measurement system (Quantum Design) for samples cut into $1 \times 2 \times 6$ mm$^3$. $T_c$ and $H_{c2}$ values were determined by 90% of superconducting transitions. In this determination, $H_{c2}$ is considered as the highest upper critical field in the samples, *i.e.*, $H_{c2}^{//ab}$, because these samples are untextured polycrystalline bulks. Note that $H_{c2}(T)$ was nonlinear; thus, the slopes of $H_{c2}(T)$ were determined according to three definitions: between 0 and 1 T (near $T_c$), 0 and 9 T (field range measured in this study), and 2 and 9 T (linear part).

Figure 1 shows the powder XRD patterns of the milled powders prior to sintering and the ground powders of sintered bulk samples with $E_{BM}$ values of 0, 50, and 590 MJ/kg. As shown in Fig. 1(a), the unsintered powders demonstrated elemental metal peaks of Fe and As at 0 MJ/kg, while Ba122 peaks were observed at 50 and 590 MJ/kg. This indicates that mechanical alloying of Ba122 occurred due to high-energy milling[29]. In the sintered bulk samples (Fig. 1 (b)), multiphase peaks of Ba122 and Fe$_2$As were observed at 0 MJ/kg, while nearly single-phase Ba122 peaks were confirmed at 50 and 590 MJ/kg. These results suggest that sintering mechanically alloyed powder is effective to obtain high purity Ba122 polycrystalline samples.

Figure 2 shows the $E_{BM}$ dependencies of (a) Co-doping level, (b and c) full-width at half maximum (FWHM) of XRD main peak of Ba122 (103) ((b) before sintering and (c) after sintering), (d) relative peak intensity of (200) and (004) to (103), (e) $a$-axis length, and (f) $c$-axis length. The Co-doping level in Fig. 2(a) shows the average value and standard deviation of the elemental analysis results for approximately 20 points on polished surface of the sintered bulk samples. The standard deviation decreases with an increase in the $E_{BM}$, which suggests that high-energy milling enhances the compositional homogeneity. There was almost no average change against $E_{BM}$, and the actual composition was $x \approx 0.084$. Microstructural observation by scanning electron microscope (not shown) demonstrated that the $E_{BM} < 40$ MJ/kg samples contained impurities such as Fe, FeAs, and Fe$_2$As, which are not detected by XRD. Since our sintering temperature is rather low (600°C), sufficient mechanical alloying is required to obtain single-phase Ba122 samples. In the following, the $E_{BM} > 40$ MJ/kg samples are treated as single-phase Ba122 and discussed.

With an increase in the $E_{BM}$, systematic broadening of FWHM was observed for the powder prior to sintering (Fig. 2 (b)). A similar but less pronounced systematic broadening was also observed for the sintered samples (Fig. 2 (c)). STEM observation[30] for the sintered samples showed that the grain size of Ba122 was refined with an increasing in $E_{BM}$. This is considered as one of the reasons for the increased FWHM, *i.e.*, degradation of crystallinity. The relative peak intensity of (200) increased systematically with an increase in the $E_{BM}$, while that of (004) was almost constant (Fig. 2 (d)). This indicates that the reflection from the *ab* plane was disturbed with an increase in the $E_{BM}$, since (200), (004), and (103) are influenced by the reflection from the *ac* plane, *ab* plane, and both crystallographic parameters, respectively. These suggest the introduction of lattice defects parallel to the *ab* plane, such as stacking faults[31], by high-energy milling. In fact, STEM observation[30] revealed linear contrasts which were parallel to each other with a spacing of several nanometers, comparable to the coherence length near $T_c$. The lattice constants of the sintered sample with $E_{BM}$ of 40 MJ/kg were $a = 3.9604(2)$Å and $c = 12.9963(7)$Å (Fig. 2 (e), (f)). The reported lattice constants of 8% Co-doped Ba122 single crystals are $a = 3.9600$[32], $3.9604(1)$[33], and $3.962$[34]Å, and $c = 12.9779$[32], $12.9793(3)$[33], and

12.991[34]Å. $a$ and $c$ of 40 MJ/kg sample are longer than those of a single crystal[32] by 0.01% and 0.14%, respectively. With an increase in the $E_{BM}$, $a$ decreased monotonically and $c$ increased to reach saturation. At 590 MJ/kg, the lattice constants were $a = 3.9588(3)$Å and $c = 13.0133(15)$Å, which means that $a$ decreased by 0.04% and $c$ increased by 0.13% from 40 MJ/kg. The observed evolution of lattice constants $a$ and $c$ cannot be explained by the change in Co-doping level because the literature states that $both$ $a$ and $c$ decrease with Co doping[32,34], and EDS analysis showed no change in the Co-doping level (Fig. 2 (a)). This opposite trend in $a$ and $c$ is frequently observed in thin films. For Ba122 thin films, it has been reported that the lattice constants and cell volume vary due to in-plane lattice strain (epitaxial strain) caused by lattice mismatch with the substrate[35,36]. Epitaxial strain can be regarded as equivalent to uniaxial pressure or tension along the $c$-axis. Figure 3 shows $\Delta c/c_0$ vs $\Delta a/a_0$ and $\Delta V/V_0$ vs $\Delta a/a_0$ for thin films with epitaxial strain[35,36] and the polycrystalline bulks examined herein, where $V$ is a cell volume, and $a_0$, $c_0$, and $V_0$ are values of the thin film deposited on the YAO substrate or values of the 40 MJ/kg sample. As can be seen, the slope $(\Delta c/c_0)/(\Delta a/a_0)$ for the polycrystalline bulks is at least three times greater in comparison to the epitaxial strain. Therefore, the defects introduced into the polycrystalline bulks by high-energy milling cannot be explained only by a lattice strain. In addition, while $\Delta V/V_0$ decreased with decreasing $\Delta a/a_0$ in the thin films, it increased in the polycrystalline bulks. In the thin films, the decrease of $\Delta V/V_0$ is due to Poisson's ratio different from 0.5[37]. In the polycrystalline bulks, the increase of $\Delta V/V_0$ suggests introduction of lattice defects with vacancy, which is consistent with the results of Fig. 2 (d)–(f) and should be responsible for the broadening of FWHM.

Figure 4 shows the temperature dependencies of the normalized electrical resistivity for the samples with $E_{BM}$ values of 50, 80, 170, and 590 MJ/kg. The data of a single crystal (Co 10%)[38] is shown for comparison. With an increase in the $E_{BM}$, the resistivity at 30 K ($\rho_{30\,K}$) increased fourfold (0.76, 2.00, 3.19, and 2.82 mΩcm, respectively), and $RRR$ ($\rho_{300\,K}/\rho_{30\,K}$) decreased by 35% (1.72, 1.64, 1.37, and 1.11, respectively). In comparison with those of Co-doped Ba122 single crystals, $\rho_{30\,K}$ is an order of magnitude higher and $RRR$ is about a half ($\rho_{30\,K} = 0.09$ (Co 6.3%)[39], 0.09 (Co 8%)[9], 0.11 (Co 10%)[40], and 0.16 (Co 10%)[38] mΩcm, $RRR = 2.69$ (Co 6.3%)[39], 1.82 (Co 8%)[9], 2.85 (Co 10%)[40], and 2.53 (Co 10%)[38]). The temperature dependencies of the electrical resistivity demonstrated metallic behavior at 30–300 K for samples with low $E_{BM}$ (< 170 MJ/kg), whereas semiconductor-like upturn at $T$ < 100 K was observed for samples with high $E_{BM}$ (> 230 MJ/kg). The inset of Fig. 4 shows the temperature dependences of the electrical resistivity under magnetic fields of 0–9 T for the samples with $E_{BM}$ values of 50 and 590 MJ/kg. Although all the samples exhibited superconducting transition and reached zero resistance even under magnetic field, they showed a broadening of resistive transition under magnetic field. Since the samples are randomly oriented polycrystalline bulks, an increase in anisotropy, refinement of grain size, and change in inter-granular structure are possible causes of the broadening. Moreover, the samples with >120 MJ/kg showed double transition. Transitions at higher and lower temperatures are considered to correspond to those of grains and grain boundaries, respectively, analogous to cuprates[41]. Such broadening or double transition is also observed in K-doped Ba122 samples[42-44]. Figures 5(a), (b), and (c) show the magnetic phase diagram, the $E_{BM}$ dependencies of (b) $T_c$, and (c) a slope of $H_{c2}(T)$. $T_c$ showed a maximum value of 26.6 K at 50 MJ/kg, which decreased slightly to 25.1 K with an increase in the $E_{BM}$. Note that the $T_c$ value was greater than optimal $T_c$ values of typical single crystals (22.0–24.8 K[4,5,9,32,38,40,45]) at all $E_{BM}$ values despite the low crystallinity of our samples. With an increase in the $E_{BM}$, $H_{c2}(T)$ near $T_c$ changed from an upward to downward curvature. Due to the curvature change near $T_c$, the slope value changed with its definition. The slope of $H_{c2}(T)$ increased with $E_{BM}$, *i.e.*, enhancement of approximately 50% from 4.1 (20 MJ/kg) to 6.2 T/K (430 MJ/kg) was observed with the most conservative definition (approximation in the linear part). Moreover, higher slope values than those of single crystals (5.0 T/K[46], 5.2 T/K[38]) and thin films (5.1 T/K[47]) were obtained at >80 MJ/kg. Anti-correlation between $T_c$ and the slope of $H_{c2}(T)$ suggests an existence of optimum intermediate $E_{BM}$ for $H_{c2}(0$ K).

The reduction in $T_c$ and increased slope of $H_{c2}(T)$ can be explained roughly by the enhanced electron scattering model induced by high-energy milling. It is considered that the introduction of lattice defects makes the samples dirtier, leading to reduction of the electron mean free path and increase of the electron scattering rate. The low temperature upturn in resistivity and change in $H_{c2}(T)$ curvature near $T_c$ suggest that the introduced lattice defects modified the electronic structure, especially the multiband structure, which strongly affects $H_{c2}$[48]. Moreover, since $c/a$ increased with an increase in the $E_{BM}$, the effective electron mass anisotropy is likely increasing. Given the relationship between the anisotropy and the effective mass, this would also increase the $H_{c2}$ anisotropy, with possible enhancement of $H_{c2}^{//ab}$. Furthermore, the noticeable increase of $H_{c2}(T)$ near $T_c$ is similar to the well-established theory of $H_{c2}$ in weakly coupled S-I-S multilayers[49]. As expected in this scenario, the stacking faults introduced by high-energy milling may behave as weakly coupled planer Josephson junctions.

Attempts have been made to introduce various types of lattice defects/strains into 122 single crystals, thin films, and polycrystalline bulks and wires. In the case of Co-doped Ba122, nonmagnetic impurities introduced by fast neutron irradiation and proton irradiation were reported to decrease $T_c$ at a ratio of $10^{-22}$ Km$^2$ and $0.5–5.8 \times 10^{-20}$ Km$^2$, respectively[16,45,47,50-55]. When epitaxial strain is introduced into thin films, $T_c$ changes continuously with $c/a$ from 16 to 28 K[35,36]. In these cases, lattice defect/strain changed $T_c$ but did not alter $H_{c2}$; therefore, the high-energy milling developed herein is a unique way to introduce lattice defects that differ from other methods and to tune the 122 system's $H_{c2}$. At the present stage, we have not been able to specifically identify the defects that correspond to the enhanced $H_{c2}$. A study into the intra-granular structure and the physical properties under higher magnetic fields should clarify the defects introduced by high-energy milling and the mechanism by which $H_{c2}$ is enhanced.

Here we briefly discuss the applicability of this defect engineering by high energy milling to the ongoing development of 122 bulks and wires. The broadening of resistive transition would cause a decrease in inter-granular $J_c$, which in turn reduces the applicable range in magnetic field. For all kinds of anisotropic, polycrystalline superconducting materials, alignment of grain orientations and/or increasing packing factor of superconducting phase are known as effective approaches to improve inter-granular $J_c$[56]. Since our samples are 65–75% in packing factor with randomly oriented grains, texturing and/or densification would be required to maximize the effectiveness of the defect engineering to $J_c$ of the bulks and wires. The introduction of anisotropic intra-granular defects and modulation of elemental pinning force in which $H_{c2}$ is a prefactor would bring new developments in flux pinning engineering of 122 polycrystalline materials.

In summary, to improve $H_{c2}$ of 122 phase IBSCs, Co-doped Ba122 polycrystalline bulk samples were synthesized as $E_{BM}$ was changed systematically, and the effects of high-energy milling on the structural and transport properties were evaluated. High-energy milling improved phase purity and introduced lattice defects into the Ba122 grains. These lattice defects deteriorated crystallinity, decreased $a$-axis length, increased $c$-axis length, and changed the shapes of $\rho(T)$ and $H_{c2}(T)$. Moreover, this resulted in continuous improvement of the slope of $H_{c2}(T)$ by 50% without changing the doping level, while $T_c$ was suppressed slightly by 5.5%. In principle, it can be expected that this method can also be applied to K-doped Ba122, P-doped Ba122, and other iron-based superconductors.

The authors would like to thank Dr. Soshi Iimura and Dr. Kota Hanzawa (Tokyo Institute of Technology) for their help with the high-field measurements. The authors also thank Dr. Kazumasa Iida (Nagoya University) and Dr. Yusuke Shimada (Tohoku University) for their fruitful discussions. This work was supported by JST CREST Grant Number JPMJCR18J4 and by JSPS KAKENHI Grants JP15H05519 and JP18H01699. AY is supported by MEXT Elements Strategy Initiative to Form Core Research Center.

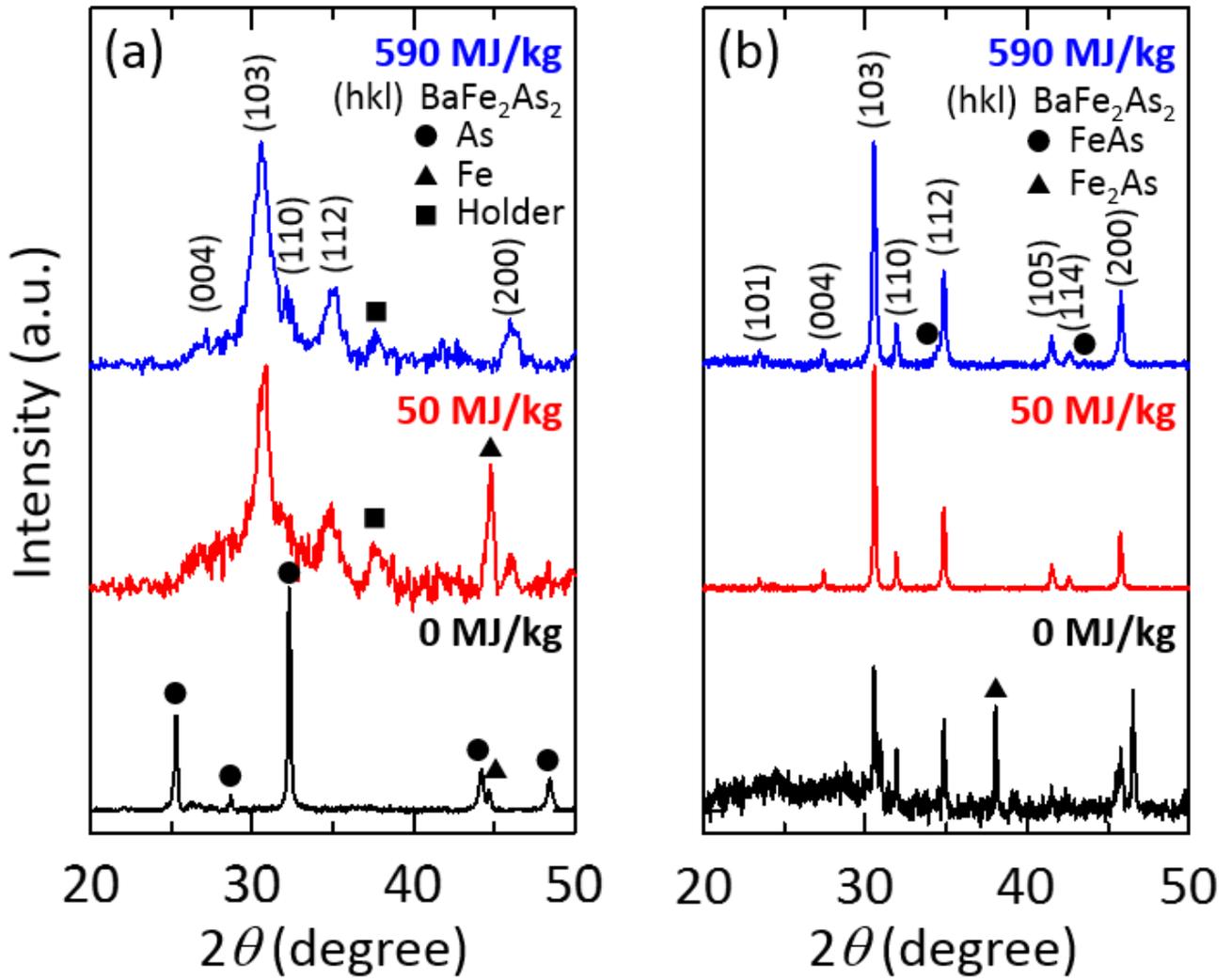

FIG. 1.

Powder XRD patterns of (a) milled powders before sintering and (b) ground powders of sintered bulk samples with $E_{BM}$ = 0, 80, and 590 MJ/kg.

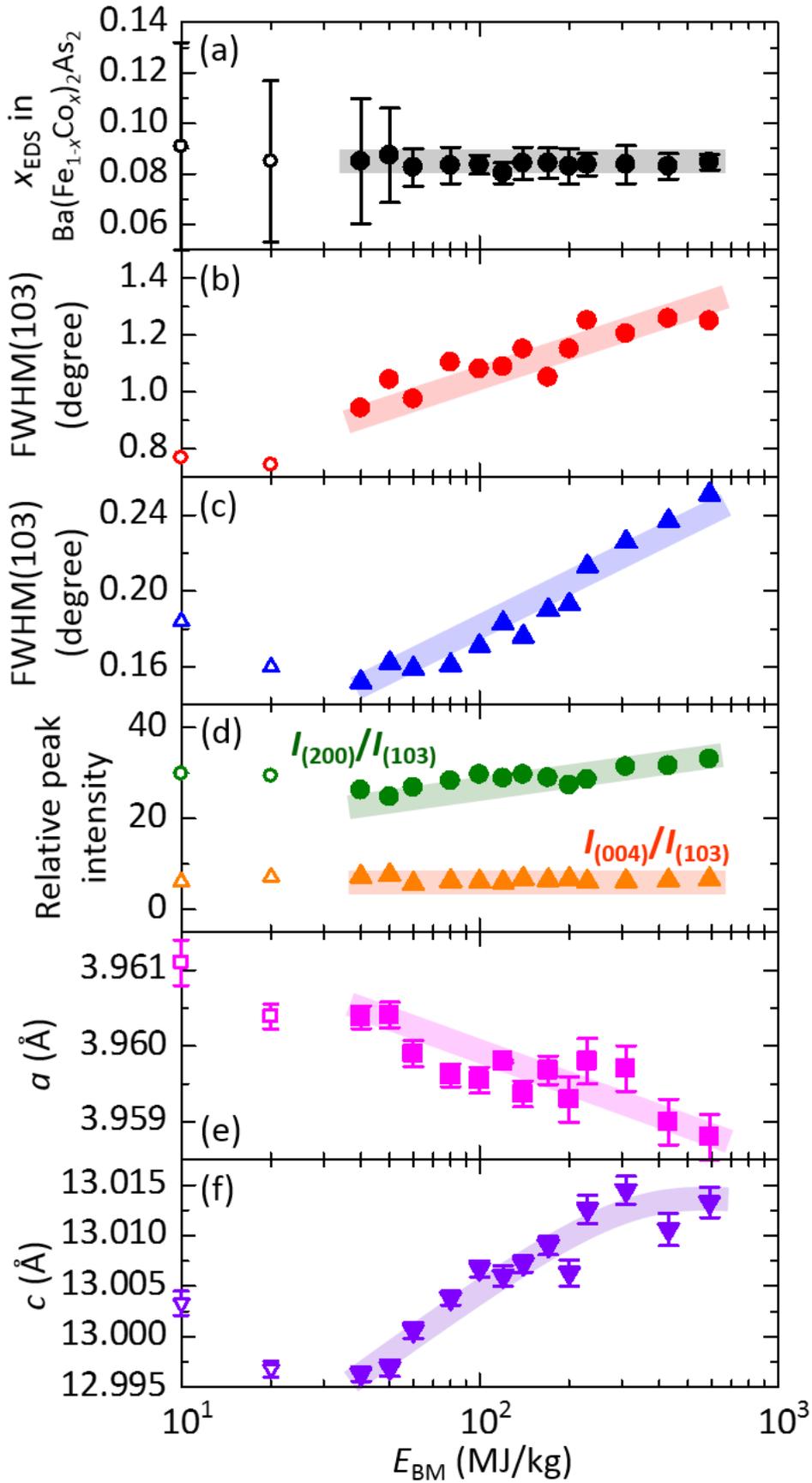

FIG. 2.

Ball-milling energy $E_{BM}$ dependences of (a) Co-doping level, (b, c) FWHM of Ba122 main peak (103), (d) relative peak intensity of (200) and (004) to (103), (e) $a$-axis length, and (f) $c$-axis length. Fig. 2 (a) is for sintered bulk samples, Fig. 2 (b) is for milled powders before sintering, and Fig. 2 (c)–(f) are for ground powders of sintered bulk samples. Open symbols are the data of multiphase samples.

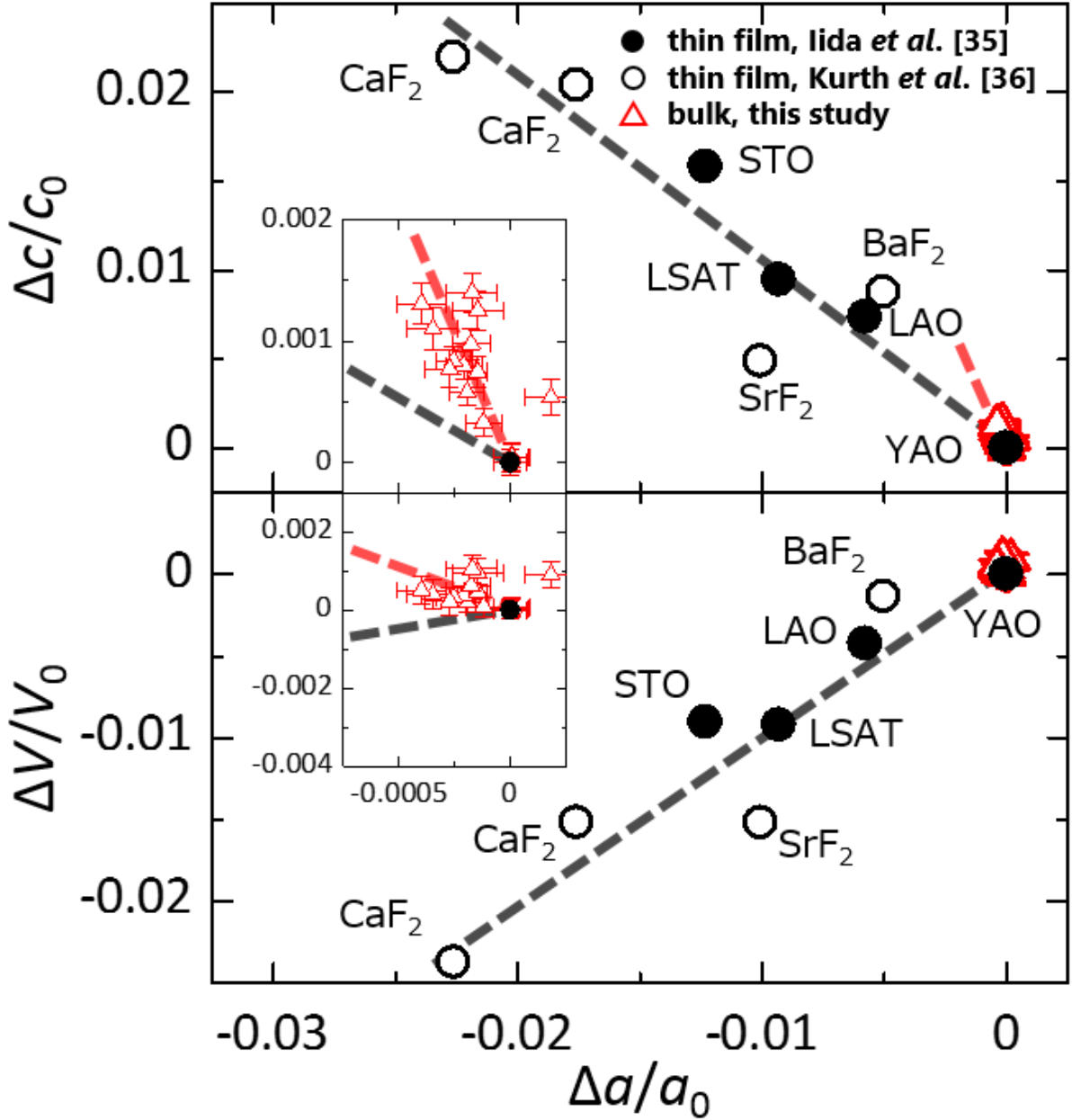

FIG. 3.

$\Delta c/c_0$ vs $\Delta a/a_0$ (upper panel) and $\Delta V/V$ vs $\Delta a/a_0$ (lower panel) for Co-doped Ba122 polycrystalline bulks (red) and Co-doped Ba122 thin films with epitaxial strain[35,36] (black). The inset shows enlarged view for polycrystalline bulks. In the case of polycrystalline bulks, $a_0$ and $c_0$ are lattice constants of $E_{BM}$ = 40 MJ/kg ($a$ = 3.9604(2)Å, $c$ = 12.9963(7)Å). In the case of thin films, $a_0$ and $c_0$ are lattice constants of the thin film on the YAO substrate ($a$ = 3.980Å, $c$ = 12.907Å).

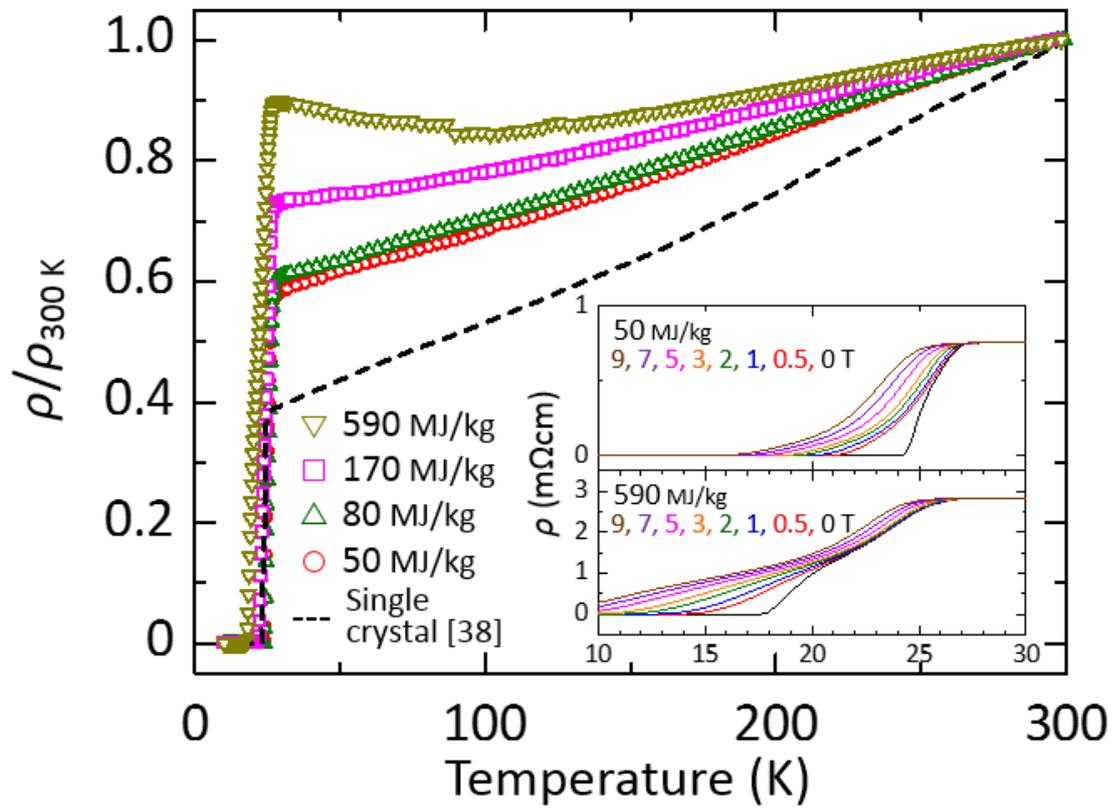

Fig. 4.

Temperature dependencies of normalized electrical resistivity for $E_{BM}$ = 50, 80, 170, and 590 MJ/kg samples. The broken line is the data of a single crystal (Co 10%)[38]. The inset shows transitions near $T_c$ (under 0–9 T) for the $E_{BM}$ of 50 and 590 MJ/kg samples.

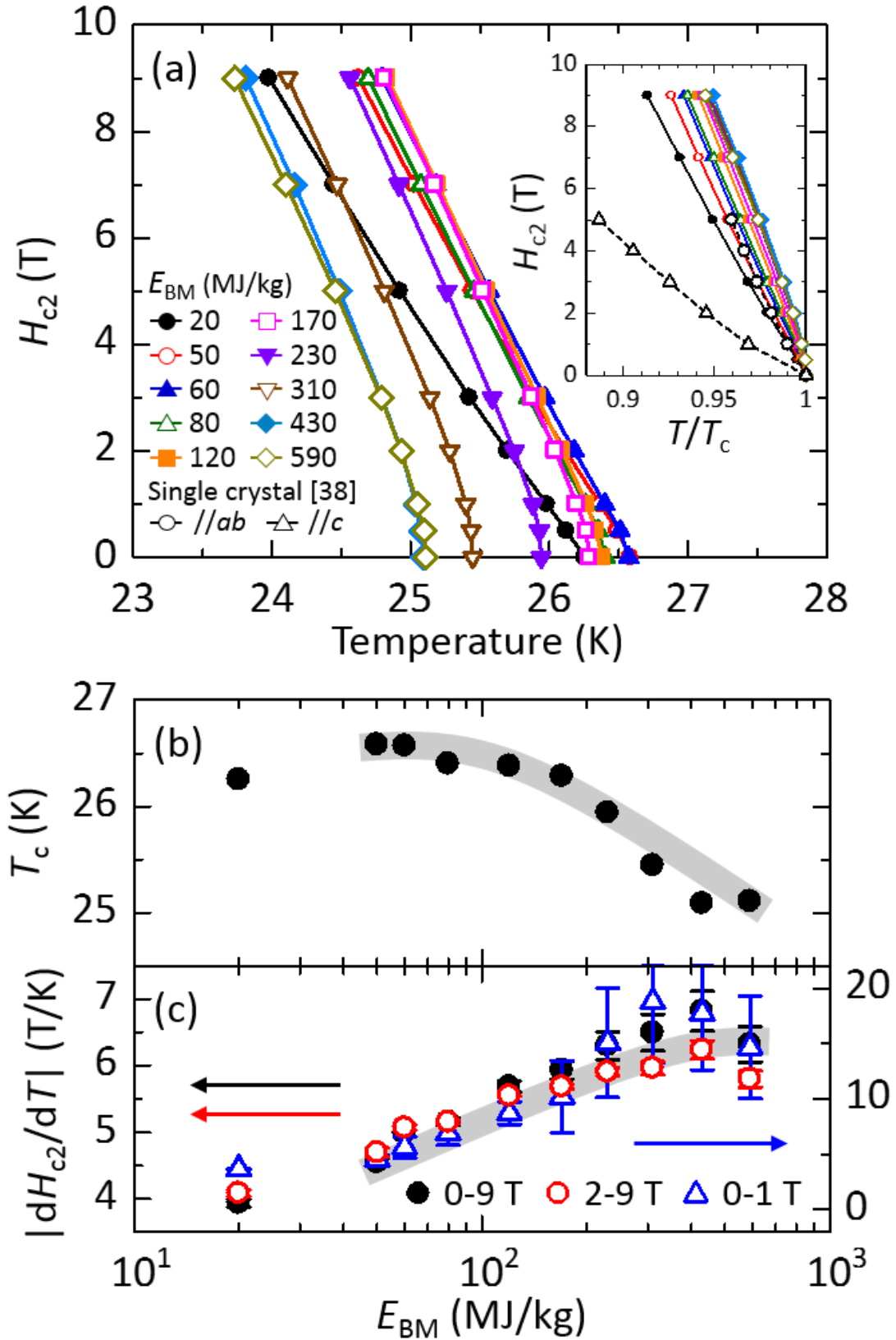

FIG. 5.

(a) Temperature dependencies of $H_{c2}$ for samples with different $E_{BM}$. In the inset, temperature is normalized by $T_c$ and the broken lines are the data of a single crystal (Co 10%)[38]. (b) Ball-milling energy ($E_{BM}$) dependence of $T_c$. (c) $E_{BM}$ dependence of slopes of $H_{c2}(T)$ between 0–1 T (near $T_c$), 0–9 T (field range measured in this study), and 2–9 T (linear part). With increasing $E_{BM}$, the slope increased by 5.1±1.6, 1.7±0.1, and 1.5±0.1 times for each range, respectively.